\documentstyle[12pt]{article}

\begin{document}
\begin{center}
{\large\bf Potential of Interplanetary Torques and Solar
Modulation
for Triggering Terrestrial Atmospheric and Lithospheric
Events}
\end{center}
\bigskip
\begin{center}
{\bf Rhodes W. Fairbridge}\\
Columbia University \& NASA/Goddard Institute for Space\\[0,2cm]
Studies, New York 10025, USA\\
{\bf G\"{o}ran Windelius}\\
Motj\"{a}rnshytten 44,
S-68391 Hagfors, Sweden\\
{\bf Hans J. Haubold}\\
UN Outer Space Office, Vienna International Centre,\\
P.O.Box 500, A-1040 Vienna, Austria
\end{center}
\bigskip
\noindent
Abstract. The Sun is forced into an orbit around the barycenter
of
the solar system because of the changing mass distributions of
the
planets. Solar-planetary-lunar dynamic relationships may form a
new
basis for understanding and predicting cyclic solar forcing
functions on the Earth's climate.\par
\clearpage
Using the NASA-Jet Propulsion Laboratory planetary ephemeris
DE-102, a time-tabel of paired synodic conjunctions has been
prepared
for the four major planets extending more than 16.000 yr in the
past and 100 yr into the future (Fairbridge and Sanders, 1987;
Sanders and Fairbridge, 1987). While all planetary orbital
periods
are somewhat irregular over
short-term intervals, when calculated over time periods $>1000$
yr,
long-term
means establish a pattern of beat frequencies that conform to a
complex harmonic system with an approximately log/log
relationship
with a constant slope. The slope formula, where v denotes
harmonic
and u denotes the period (in years), is as follows:
$$log(v) = 6.04 - log(u).$$
Without exception, all beat frequencies and resonance periods are
commensurable (in simple, whole-integer ratios), subject to
``errors'' that decrease systematically over greater time
intervals
$(10^{-5} \mbox{at} >10^4\, \mbox{yr}).$ For the lunar planets,
comparable pair beat frequencies are found in the spectral
analysis
of monthly sunspot numbers (Currie, 1973) and in the time domain
the
numbers have been found to correspond to the ephemeris
predictions
(Verma, 1986).\par
This astronomical time-table provides a basis for comparison with
the
history of behavior patterns of (Fairbridge, 1989; Shirley,
Sperber, and Fairbridge, 1990; Haubold and Beer, 1992):\par
(a) {\it The Sun:} monitored over $> 90.000$ yr by use of
$^{18}O$
isotopes in ice cores and deep-sea deposits; over nearly 10.000
yr
by use of $^{14}C$ and $^{10}Be$ isotopes in tree rings and ice
cores; over $> 2000$ yr by observations of auroras and sporadic
sunspot events, over $> 300$ yr by telescopic sunspot data; and
during
the 20th century by multidisciplinary sensing systems, most
recently from satellite platforms.\par
(b) {\it The Earth:} monitored over $> 4$ billion years through
the
geological record (which, through evidence of mineral equilibria,
sedimentological mechanisms and the uninterrupted continuum of
organic evolution, conclusively proves the fundamental stability
of
the Solar System); monitored over the 560 million year
Phanerozoic
interval by analysis of sequence stratigraphy (suggesting the
stability of the Earth/Moon orbital relationships in the 20, 41,
93
kyr ``Milankovitch'' cyclicity).\par
Calibration of planetary/Sun/Earth relationships has been
obtained
using the chronometry of the integrated tree ring record
(accurate
to $\pm 1$ yr), and supplemented by ice core data, both of which
furnish isotopic data reflecting (1) Sun's emissions, (2) Earth's
spin rate, axis parameters, and geomagnetic variables. Both, in
turn apply forcing to the Earth's climatic dynamics.\par
Planetary triggering of dynamic events on the Sun and Earth is
believed to operate primarily through exchange of interplanetary
torques. Although the Saturn-Jupiter pair (synodic lap: 19.859
yr)
carry 86\% of the angular momentum of the Solar System, it is the
synod of the two outermost giants (Neptune-Uranus: lap 171.4095
yr)
that acts as the primary trigger to a sporadic ripple effect that
propagates through the system. Other pairs (USL, SJL, EVL, etc.)
set up secondary ripples. The closest conjunctions are extremely
rare, but create a serial domino effect through the system.\par
On planet Earth two `` control dates'' have been selected that
help
provide a framework for constructing a climatic chronology. For
warm cycles the best data is -1090 (sidereal) or 3040 B.P.
(before
A.D. 1950), principal fluctuations occur at multiples of 171.4 yr
(NUL $\times$3, 9, 27, etc.). For the cold cycles the date is
-3321
or 5271 B.P., together with the principal multiples (at NUL
$\times$
2, 6, 12, etc.).\par
It is now postulated that this knowledge of long-term historical
behavior constitutes a powerful tool for estimating future
hazardous events on planet Earth, e.g. seismic crescendos,
volcanic
eruptions, climatic crises. The year AD 1993 was long recognized
a
an NUL synod and closely clustered ($\pm$5 yr) with other pairs.
The 171.4 yr cycle is $\pm$ 1.4 yr; 1993 is 18 $\times$ NUL, +2
yr, after -1090. It did in fact turn out to be an ``interesting''
year. A much more refined assessment of astronomical triggering
of
terrestrial hazard potentials should now be possible.\par
\clearpage
\noindent
\begin{center}
{\bf Acknowledgements}
\end{center}
The NASA-Goddard Institute for Space
Studies (New York City) has materially aided this research in
many
ways, also assisted by an NSF (Engineering) grant (R.W.F.,
1988/90).\par
\medskip
\noindent
\begin{center}
{\bf References}
\end{center}
\noindent
Currie, R.G.: 1973, Astrophys. Space Sci. \underline{20},
509.\par
\medskip
\noindent
Fairbridge, R.W.: 1989, Quaternary International \underline{2},
83.\par
\medskip
\noindent
Fairbridge, R.W. and Sanders, J.E.: 1987, in M.R. Rampino, J.E.
Sanders,\par
W.S. Newman, and L.K. K\"{o}nigsson (eds.), 'Climate:
History,\par
Periodicity, Predictability', Van Nostrand Reinhold, New
York,\par
pp. 446-471,\par
\medskip
\noindent
Haubold, H.J. and Beer, J.: 1992, in W. Schr\"{o}der and J.P.
Legrand (eds.),\par
'Solar-Terrestrial Variability and Global Change',
Selected papers\par
from the symposium of the Interdivisional Commission on
History\par
(ICH) of the International Association of Geomagnetism and\par
Aeronomy (IAGA), XX General Assembly of IUGG, Vienna,
Austria,\par
11-24 August 1991, pp. 11-34.\par
\medskip
\noindent
Sanders, J.E. and Fairbridge, R.W.: 1987, in M.R. Rampino,\par
J.E. Sanders, W.S. Newman, and L.K. K\"{o}nigsson (eds.),\par
'Climate: History, Periodicity, Predictability', Van Nostrand\par
Reinhold, New York, pp. 475-541.\par
\medskip
\noindent
Shirley, J.H., Sperber, K.R., and Fairbridge, R.W.: 1990, Solar
Phys.\par
\underline{127}, 379.
\end{document}